\documentclass[twocolumn,floatfix,longbibliography]{revtex4}
\usepackage{graphics,epsfig,amsfonts,amssymb,amsmath,xcolor,ulem}

\begin{document}
\title{The nature of correlations in the insulating states of twisted bilayer graphene}
\author{J.M. Pizarro}
\author{M.J. Calder\'on}
\author{E. Bascones}
\email{leni.bascones@icmm.csic.es}
\affiliation{Materials Science Factory. Instituto de Ciencia de Materiales de Madrid (ICMM). Consejo Superior de Investigaciones Cient\'ificas (CSIC), Sor Juana In\'es de la Cruz 3, 28049 Madrid (Spain).}

\date{\today}
\begin{abstract}  
The recently observed superconductivity in twisted bilayer graphene emerges from insulating states believed to arise from electronic correlations. While there have been many proposals to explain the insulating behaviour, the commensurability at which these states appear suggests that they are Mott insulators. 
Here we focus on the insulating states with $\pm 2$ electrons or holes with respect to the charge neutrality point. We show that the theoretical expectations for the Mott insulating states are not compatible with the experimentally observed dependence on temperature and magnetic field if, as frequently assumed, only the correlations between electrons on the same site are included. We argue that the inclusion of non-local (inter-site) correlations in the treatment of the Hubbard model can bring the predictions for the magnetic and temperature dependencies of the Mott transition to an agreement with experiments and have consequences for the critical interactions, the size of the gap, and possible pseudogap physics. The importance of the inter-site correlations to explain the experimental observations indicates that the observed insulating gap is not the one between the Hubbard bands and that antiferromagnetic-like correlations play a key role in the Mott transition.
\end{abstract} 
\maketitle

\section{Introduction}
Unexpected insulating states have been recently observed upon doping a graphene bilayer with a small twist angle $\sim1.08^o-1.16^o$~\cite{caoNat2018_1}. The stacking misorientation creates a moir\'e pattern 
with a superlattice modulation corresponding to thousands of atoms per unit cell. The insulating states, believed to be due to electronic correlations~\cite{caoNat2018_1}, were originally observed when the charge per moir\'e cell is $\pm 2$ with respect to the charge 
neutrality point (CNP). More recently, insulating states when the twisted bilayer graphene (TBG) is doped with 1 or 3 carriers have also been observed~\cite{yankowitz-arXiv2018,jarillobeijing}. 

The experimental characterization of the insulating states with $\pm 2$ electrons or holes, in which we focus from now on, reveals that the system becomes metallic with increasing temperature and in a magnetic field~\cite{caoNat2018_1}. In addition, 
there is no sign of symmetry breaking due to magnetic order. 
Superconductivity emerges from the insulating statel~\cite{caoNat2018_2,yankowitz-arXiv2018,jarillobeijing}. Understanding the nature of the insulating states is key to uncover the origin of the superconductivity. 

The band structure of TBG at the charge neutrality point features
two doubly-degenerate Dirac cones~\cite{dossantosPRL2007}. The nearby bands at both positive and negative energies become very narrow at the experimental twist angle~\cite{bistritzerPNAS2011}.  
The electronic charge in these narrow bands is believed to be concentrated at the center of each moir\'e cell, in the so-called AA 
regions which form a triangular superlattice~\cite{LaissardiereNanoLett2010}. 

Several authors have proposed different origins for the insulating states, many of them relying on specific properties of the density of states at the Fermi level~\cite{yuanPRB2018,zhang-arXiv2018,poPRX2018,liu-arXiv2018,roy-arXiv2018,isobe-arXiv2018,dodaroPRB2018,koshinoPRX2018}. However,  an important clue comes from the fact that the fillings at which these states are experimentally found correspond to an integer number of electrons/holes per moir\'e unit cell, as for Mott states in Hubbard models.  In fact, the exact commensurability has been now observed for a larger range of twist angles~\cite{caoNat2018_1,yankowitz-arXiv2018,jarillobeijing}, for which considerably different density of states are expected. The proximity of a metallic gate, at a distance of only $1-2$ moir\'e lattice constants from the TBG~\cite{caoNat2018_1}, is expected to screen the long-range character of the Coulomb interaction.  For all these, an effective Hubbard model for the moir\'e superlattice seems a plausible starting point to describe the insulating and superconducting states in a TBG. Due to the valley degree of freedom the effective Hubbard model should include at least two orbitals~\cite{caoNat2018_1}.

In contrast to single orbital systems, for which the insulating Mott state appears only at half-filling (one electron per site in the single orbital case), in multi-orbital Hubbard models Mott states are also found for other fillings with an integer number of electrons per site\cite{rozenbergPRB1997}. At large interactions the energy of the system is reduced if an integer number of electrons is localized at each atomic site and the system goes through a metal-insulator Mott transition at a critical on-site interaction  $U^c$ which depends on the filling and on other interaction parameters such as Hund's coupling $J_H$. Though localized electrons have a tendency to order magnetically, the Mott transition does not require any symmetry breaking. For interactions smaller than the critical interaction for the Mott transition, a metallic correlated state is found with the spectral weight partially redistributed to many-body Hubbard bands. 

Local correlations, namely correlations between electrons on the same site, induce a Mott transition at integer fillings via the opening of a charge gap. Approaches usually used to address the Mott transition, such as single-site Dynamical Mean Field Theory (DMFT)~\cite{georgesRMP1996} or slave particle techniques~\cite{KotliarPRL1986,FlorensPRB2004,demediciPRB2005,yuPRB2012}, only include these on-site correlations.  However, even if in the Hubbard model interactions are restricted to electrons on the same site, inter-site (non-local) correlations are established. The interplay between these magnetic and orbital non-local correlations and the charge degree of freedom affects significantly the Mott transition. Here we show that experimental observations on the insulating states in TBG are not consistent with expectations including only local correlations. We support our claim by studying their effect
on the two-orbital Hubbard model on the honeycomb lattice for the moir\'e superstructure with a U(1) single-site slave-spin mean-field technique~\cite{yuPRB2012}. We argue that the inclusion of non-local correlations can heal this disagreement, indicating that antiferromagnetic-like correlations play a key role in the localization of the carriers.

\section{Model}
The hamiltonian for the moir\'e superlattice comprises the kinetic energy $H_t$ and the on-site interaction term $H_U$. The tight-binding models for the moir\'e superlattice are not defined in the basis of the atomic $p_z$ orbitals of the carbon atoms, but on effective orbitals from Wannier projections for the low energy bands. These effective orbitals are not defined in a particular graphene layer but in the TBG. However,  the most suitable model is still under discussion~\cite{yuanPRB2018,zhang-arXiv2018,poPRX2018,roy-arXiv2018,kangPRX2018,isobe-arXiv2018,dodaroPRB2018,rademaker-arXiv2018,koshinoPRX2018,ZouPRB2018,Po-arXiv2018-2,Angeli-arXiv2018}. Although the AA regions form a triangular lattice, tight binding models with such symmetry spanning only the flat bands do not preserve the Dirac cones~\cite{poPRX2018}. Tight-binding models for different lattices and number of bands have been proposed. 

Here we focus on the simplest proposed model which reproduces the Dirac cones: a two-orbital tight-binding hamiltonian on the honeycomb lattice~\cite{yuanPRB2018,poPRX2018,zhang-arXiv2018,kangPRX2018,liu-arXiv2018} with orbitals presumably centered on the AB and  BA regions featuring lobes directed toward the AA stacking centers. 

The energy bands of the two-orbital honeycomb model~\cite{yuanPRB2018,poPRX2018,zhang-arXiv2018,kangPRX2018,liu-arXiv2018} mimic the flat bands around the charge neutrality point with bandwidth $W\sim10$ meV~\cite{bistritzerPNAS2011}. There are two bands with energy 
$E>0$ and two bands with energy $E<0$ originating in the valley degree of freedom of each of the graphene layers.   For simplicity, we only include the intraorbital hopping $t$ to first nearest neighbors ($t\sim 2$ meV). Although this assumption makes the model particle-hole symmetric with respect to half-filling, 
see the inset of Fig.~\ref{fig:Fignofield} for the density of states, it does not affect our conclusions. Below, all energies are in units of $t$. 

The filling per site in the honeycomb lattice is defined as $x=n/2N$ with  $n$ ($N$) the number of electrons (orbitals) per site. Note that we define the filling from the bottom of the bands, not from the charge neutrality point as in the original experimental papers~\cite{caoNat2018_1,caoNat2018_2}. In our notation, the filling at the Dirac points (the charge neutrality point of the TBG) corresponds to half-filling $x=1/2$ (on average 2 electrons per site,  one per orbital, in the honeycomb 
lattice, 4 per moir\' e cell). Experimentally, at this filling the system has the semi-metallic character expected from the Dirac cone dispersion and the density of states vanishes, see the inset of Fig.~\ref{fig:Fignofield}. At $x=1/4$ $(n=1)$ and $x=3/4$ $(n=3)$,  the density corresponding to the $\mp 2$ electrons with respect to the CNP, the Fermi surface is centered around $\Gamma$ 
and the density of states does not show any special features.

For multi-orbital systems, the interaction term $H_U$ includes the intra- and inter-orbital interactions $U$ and $U'$, the Hund's coupling $J_H$, and pair-hopping $J'$ terms~\cite{castellaniPRB1978,yuanPRB2018}: 
\begin{eqnarray}
\nonumber
  &H_U&   = 
   U\sum_{j\gamma}n_{j\gamma\uparrow}n_{j\gamma\downarrow}
 +  (U'-\frac{J_H}{2})\sum_{j\{\gamma>\beta\}\sigma\tilde{\sigma}}n_{j\gamma\sigma}n_{j\beta\tilde{\sigma}}
\\ 
& -&  2J_H\sum_{j\{\gamma >\beta\}}\vec{S}_{j\gamma}\vec{S}_{j\beta}
 +   J'\sum_{j\{\gamma\neq
  \beta\}}c^\dagger_{j\gamma\uparrow}c^\dagger_{j\gamma\downarrow}c_{j\beta\downarrow}c_{j\beta\uparrow}
 \,
\label{eq:hamiltoniano}
\end{eqnarray}
with $\gamma$ and $\beta$ labelling the orbitals, $j$ the sites and $\sigma, \tilde \sigma$ the spin.
Here $J'=J_H$ and $U'=U-2J_H$~\cite{yuanPRB2018,castellaniPRB1978}. $U$ has been estimated to be $\sim 20$ meV~\cite{caoNat2018_1}. The value of the Hund's coupling is not known. Yuan and Fu~\cite{yuanPRB2018} 
assume  $J_H=0$ and $U=U'$. On the other hand, Dodaro {\it et al}~\cite{dodaroPRB2018} argue that once the effect of the phonons of the TBG with frequencies 
$\omega \sim 200$ meV is included the effective $J_H$ becomes negative, as for alkali-doped fullerides~\cite{nomuraJPCM2016}. 
We do not make any a priori assumption on the value or the sign of $J_H$.  We assume the equality $U'=U-2J_H$ remains valid for $J_H<0$~\cite{nomuraJPCM2016}. 

We analyze this model with a single site U(1) slave-spin mean-field~\cite{yuPRB2012} technique. In the slave-spin approach 
the electron operator is written in terms of slave-spin and auxiliary fermion operators.  This procedure enlarges the Hilbert space and the unphysical states are eliminated requiring the system to satisfy a certain constraint. 
In the single-site approximation the 
operators in different lattice sites are assumed to be uncorrelated, namely only local correlations are included. The pair-hopping and the spin-flip terms are dropped in the calculation. We 
refer the reader to Refs.~\cite{demediciPRB2005,yuPRB2012}, where the technique is explained with great detail.

The slave-spin approach has been extensively used to study electron correlations and Mott transitions in multi-orbital systems, such as iron superconductors, Hund metals and non-degenerate systems with orbital selective Mott physics~\cite{demediciPRB2011,yuPRB2012,yiPRL2013,demediciPRL2014,CalderonPRB2014,FanfarilloPRB2015,FanfarilloPRB2017}. 
In particular, in multiorbital systems, slave-spin 
approaches~\cite{demediciPRB2005,demediciPRB2011,yuPRB2012,yiPRL2013,demediciPRL2014,CalderonPRB2014,FanfarilloPRB2015,FanfarilloPRB2017}, in Z$_2$~\cite{demediciPRB2005}  or U(1)~\cite{yuPRB2012} versions, have proven 
to be very useful to calculate the critical value of the interaction $U^c_x$. The predictions from Dynamical Mean Field Theory (DMFT) and those from slave-spin  are compared in Ref.~\cite{FanfarilloPRB2015}  for the quasiparticle weight and the critical interaction for several electronic fillings and values of the Hund's coupling. Very good agreement between both techniques is found~\cite{FanfarilloPRB2015,footnote-DMFT-slavespin}. 
The slave-spin technique does not give information on the spectral weight redistribution which takes place with increasing interactions:
Spectral weight is transferred from the quasiparticle peak to the incoherent/many body Hubbard bands, producing a three peak spectral function.  We obtain this information from previous calculations performed with DMFT~\cite{georgesRMP1996} in the literature and cited accordingly in the text.

\section{Comparison with experimental observations}

\subsection{Doping dependent critical interaction}
\begin{figure}
\includegraphics[clip,width=0.45\textwidth]{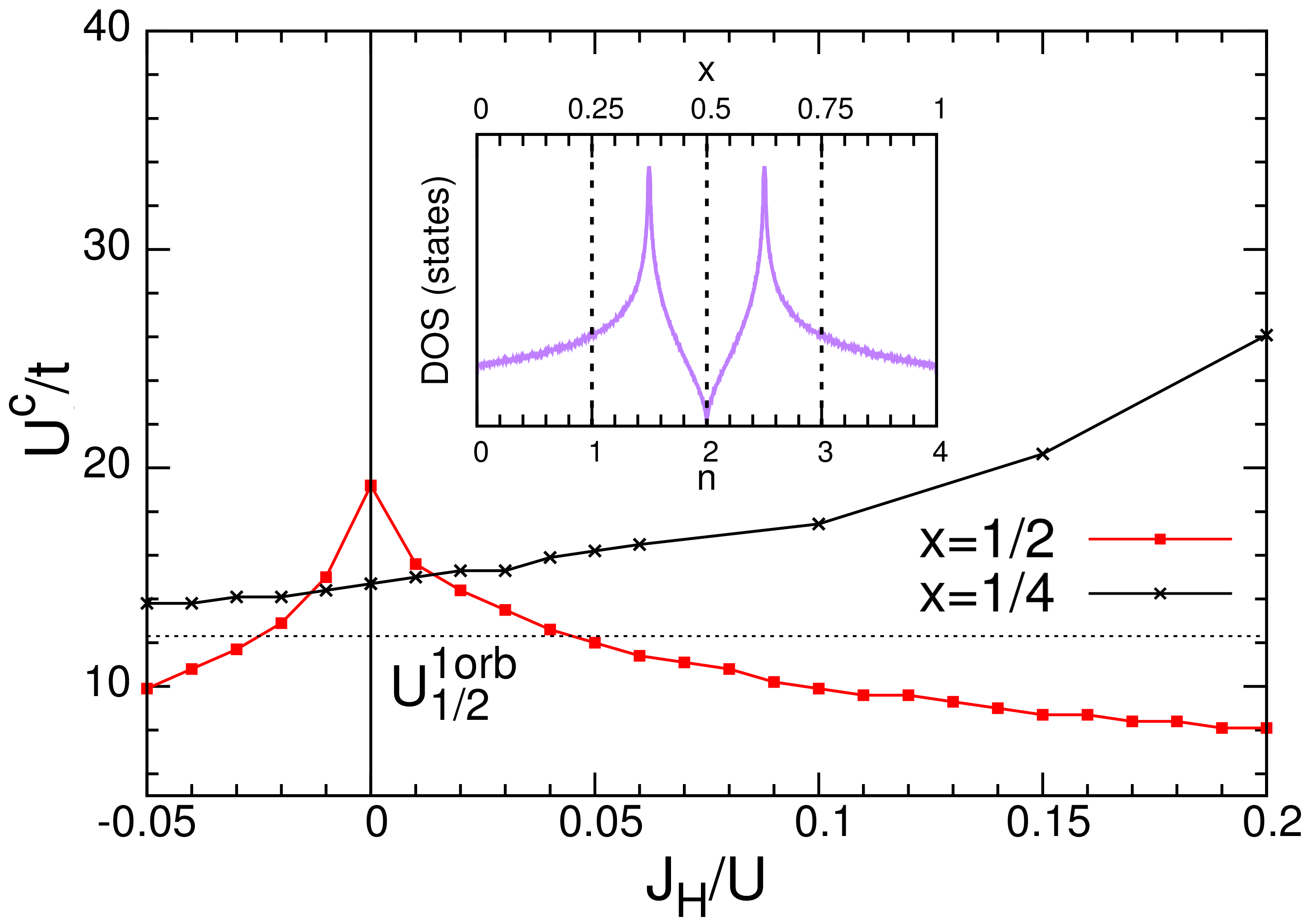}
\caption{
Interaction $U^c_x$ at which the insulating Mott state emerges for quarter- ($x=1/4$) and half-filling ($x=1/2$) as a function of Hund's coupling $J_H$ in the two-orbital honeycomb lattice. $U^c_{1/4}<U^c_{1/2}$  is satisfied only in a very small range of $J_H$ centered around $J_H\approx 0$ for which $U^c_{1/4}=14.7t$ and $U^c_{1/2}=19.2t$.  The dotted line corresponds to $U_{1/2}^{1\rm orb}$, the critical interaction for the Mott transition in a single-orbital model in the honeycomb lattice. The inset shows the non-interacting density of states as a function of electronic filling. The lower horizontal axis shows the number of electrons 
$n$ per lattice site in the two orbitals and the upper axis shows the electronic filling with respect to its maximum value. Half-filling $x=1/2$ corresponds to the charge neutrality point. Dashed lines mark 
the fillings at which the Mott insulating states, corresponding to $\pm 2$ electrons from the CNP, are found. Note that the van Hove singularities are found at different fillings. } 
\label{fig:Fignofield} 
\end{figure}

In a two-orbital model the insulating states can be found at fillings $x=1/4, 1/2$ and $3/4$. 
The insulating states with $\pm 2$ electrons with respect to the CNP are found at $x=1/4$ and $x=3/4$, with 1 and 3 electrons per site (2 and 6 per moir\'e unit cell), respectively, with respect to the bottom of the band. Due to the particle-hole symmetry of the model, below we restrict the discussion to $x \leq 1/2$. 

Agreement with the experiment requires that $U^c_{1/4}<U^c_{1/2}$, and that the interaction $U$ is intermediate between these two values, such that the system is an insulator at quarter-filling and a metal at half-filling. $U^c_x$ depends on the charging energy cost $\Delta^U_x=E(n+1)-E(n-1)-2E(n)$, in competition with the kinetic energy gain.  Here $E(n)$ is the interaction energy for $n$ electrons in the atom. $\Delta^U_x$ depends on the interaction $U$, on the Hund's coupling $J_H$ and on the filling $x$, see Appendix~\ref{app:Uc}. 
The larger the $\Delta^U_x$, the smaller the $U^c_x$, i.e. the system becomes a Mott insulator at smaller values of the interaction.  

For most of the values of $J_H$, positive or negative, $\Delta^U_{1/2}>\Delta^U_{1/4}$, leading to $U^c_{1/2}<U^c_{1/4}$, in disagreement with the experimental observations. At zero Hund's coupling 
$\Delta^U_{1/2}=\Delta^U_{1/4}=U$.  On spite of having the same charging energy cost, as seen in Fig.~\ref{fig:Fignofield}  and previously discussed~\cite{HanPRB1998,rozenbergPRB1997,FlorensPRB2004,yuanPRB2018} for small values of $J_H$, orbital fluctuations stabilise the metallic state up to larger values of $U$ at half-filling than at quarter-filling, such that $U^c_{1/2}>U^c_{1/4}$, consistent with experiments~\cite{yuanPRB2018}. 
From now on, we focus on the small range of Hund's coupling $-0.01<J_H/U<0.01$ in which the condition $U^c_{1/4}<U^c_{1/2}$ is satisfied. We approximate such small values by $J_H=0$. We note that due to the larger degeneracy of the multi-orbital insulating states, these orbital fluctuations make both $U^c_{1/4}$ and $U^c_{1/2}$ at small $J_H$ larger than $U^{1 \rm orb}_{1/2}$, the critical interaction for the Mott transition in the single-orbital model in the honeycomb lattice.

\begin{figure}
\leavevmode
\includegraphics[clip,width=0.45\textwidth]{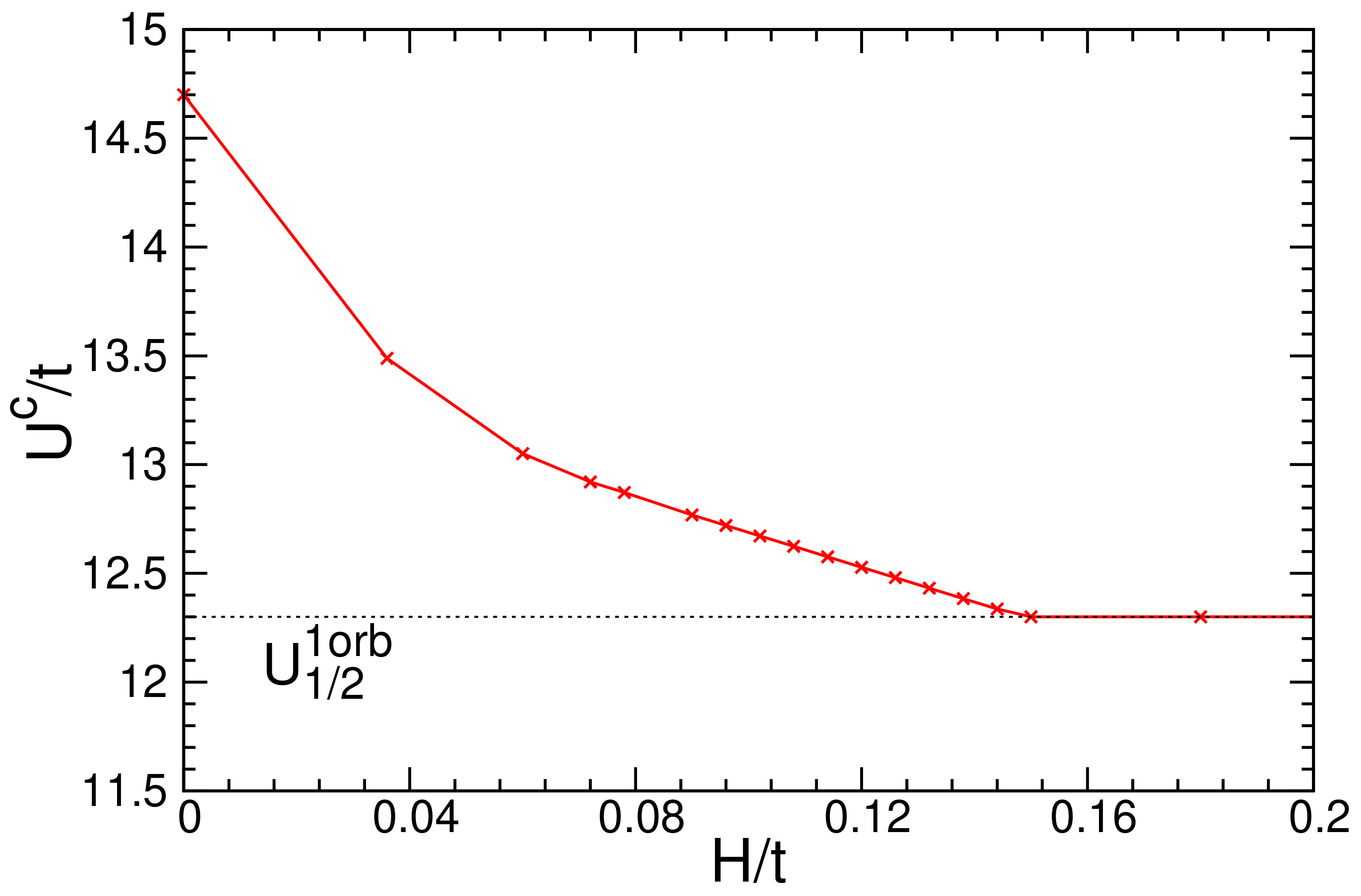}
\caption{
 $U^c_{1/4}$ as a function of the magnetic field for $J_H=0$. $U^c_{1/4}$ decreases with increasing field due to the suppression of orbital fluctuations. At large fields the critical interaction saturates to the value expected in a single-orbital model $U^{1\rm orb}_{1/2}$, see text.   
}  
\label{fig:Figfield} 
\end{figure}

\subsection{Magnetic field dependence}
Experimentally, the insulating state at $x=1/4$ becomes metallic in a magnetic field $H\sim3-6$T ($\mu_BH\sim 0.18-0.36$ meV)~\cite{caoNat2018_1}. This behaviour does not depend on whether the field is applied perpendicular or parallel to the TBG sample, 
suggesting that the suppression of the insulating state is due to the Zeeman effect. This is consistent with the fact that for localized electrons the orbital effect is not expected to play a role.   

Agreement with experiments requires that the critical interaction for the Mott transition $U^c_{1/4}$ increases in a magnetic field. However, contrary to the experimental observations, we find that in the local approximation a magnetic field decreases $U^c_{1/4}$ promoting the insulating tendencies, see Fig.~\ref{fig:Figfield}. The effect of the magnetic field $H$ is introduced via the Zeeman term $H\sum_{j\gamma} (n_{j\gamma\uparrow}-n_{j\gamma\downarrow})$.  At $x=1/4$, with $J_H=0$, the atomic gap $\Delta^U_{1/4}=U$ does not depend on $H$. When a magnetic field is applied, the Zeeman term favors spin polarization. When the bands are completely 
spin polarized (for instance, the spin up band is emptied) the behaviour of the spin down band becomes equivalent to that of a single orbital model at half-filling.
As discussed above, the critical interaction for a single orbital system $U^{1\rm orb}_{1/2}$ is smaller than $U^c_{1/4}$ at $H=0$. The magnetic field lifts the spin degeneracy suppressing the orbital fluctuations. As a consequence, $U^c_{1/4}$ decreases with the magnetic field towards $U^{1\rm orb}_{1/2}$. This is shown in Fig.~\ref{fig:Figfield}.  
More details are given in the Appendix~\ref{app:Uc}.

\subsection{Temperature dependence and gap size} 
\begin{figure*}
\leavevmode
\includegraphics[clip,width=0.6\textwidth]{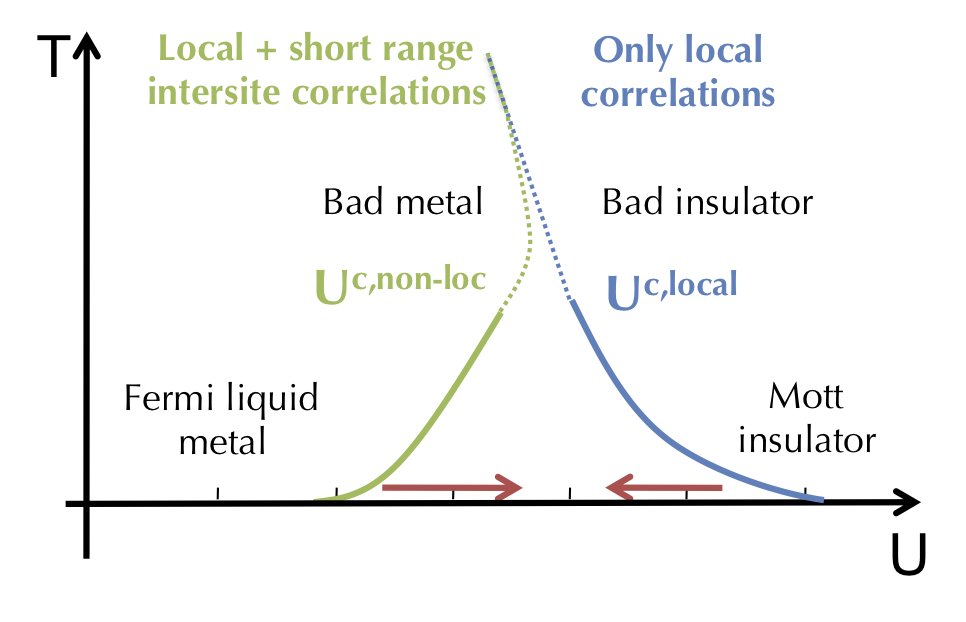}
\caption{
Sketch of the temperature-interaction phase diagram for the Mott transition in the paramagnetic single-orbital case, previously obtained in DMFT (only on-site (local) correlations are included, blue lines at the right) and CDMFT (inter-site (non-local) correlations are included, green lines at the left).~\cite{georgesRMP1996, parkPRL2008}. Above a given temperature the phase transition turns into a crossover (dashed lines). With only on-site correlations, the insulating behaviour is enhanced when the temperature is increased. For $U$ below the blue line, the metallic state turns into an insulator with increasing temperature. Above the blue line, the system is always an insulator and its resistivity decreases with increasing temperature. When short-range inter-site correlations are included, the critical interaction decreases (green line) and the low $T$ slope reverses due to the short-range singlet formation. Non-monotonic behaviour is observed and at sufficiently high-temperature both local and non-local crossover lines converge into a single one. In the region of parameters between the local and non-local lines, the system is metallic if only local correlations are considered, and insulating with inter-site correlations. In the later case, the system is insulating below the green line and metallic above it. The red arrows indicate the effect of a magnetic field on the critical interactions.
}  
\label{fig:Figtemp} 
\end{figure*}

The insulator to metal transition observed 
experimentally with increasing temperature $T$ is also at odds with the behaviour expected from a charge gap emerging from local correlations.  
In single-orbital Hubbard models, the $T$ dependence of the Mott transition in the paramagnetic phase has been extensively studied using single-site DMFT~\cite{georgesRMP1996,TerletskaPRB2013}. The results are summarized in Fig.~\ref{fig:Figtemp} (see the blue line on the right) where $U^{c,\rm local}$ refers to $U^{1\rm orb}_{1/2}$.  If at low $T$ the system is insulating (for $U>U^{\rm c,local}$) the resistivity decreases with increasing $T$, but the metallic behaviour is not restored~\cite{TerletskaPRB2013}. 
Below $U^{c,\rm local}$ the system is metallic and at a critical $T$ there is a first order transition, blue solid line, to an insulating state.  This happens because in the local approximation the entropy in the Mott state is larger than in the metallic one~\cite{fazekas-book,georgesRMP1996}.   This phenomenology has been experimentally observed in oxides~\cite{LimeletteScience2003} and remains unchanged in the two-orbital case~\cite{rozenbergPRB1997}. Hence, if only local correlations are included, the insulating behaviour is not suppressed with $T$, contrary to the experimental observations.  

Previous works have pointed out that the observed small gap $\sim 0.3$ meV is also in contradiction with the expected gap in a Mott insulator~\cite{dodaroPRB2018,liu-arXiv2018}. In this statement it is implicitly assumed that the insulating gap equals the gap between the Hubbard bands $\sim U-W \sim (U, W)$, with $W$ the bandwidth, as obtained by the single-site DMFT treatment of the Hubbard model~\cite{georgesRMP1996}. 

The lack of agreement between the experimental observations (dependence on magnetic field and $T$, and the gap size) and theoretical predictions including only local correlations is restricted neither to the two-orbital character of the Hubbard model nor to particular lattices~\cite{footnote-field1orb} or tight-binding models. 
These disagreements indicate that the experimentally observed insulating state cannot be described as a Mott state considering only local correlations. 
Below we argue that the inclusion of inter-site correlations in the Mott transitions could be the clue to reconcile theory and experiment.

\section{Effect of inter-site correlations} 
Even within the Hubbard model (with only on-site interactions) inter-site correlations are generated. The best known example is the single orbital case at half-filling for which antiferromagnetic (Heisenberg-like) correlations between the local electrons in neighboring sites emerge. In multi-orbital systems the inter-site correlations involve both magnetic and orbital degrees of freedom, and depend on $J_H$~\cite{NomuraPRB2014,NomuraPRB2015} and the filling.  For a two-orbital system with $x=1/4$ positive (negative) Hund's coupling promotes ferromagnetic and antiferroorbital (AFM and ferroorbital) correlations~\cite{momoiPRB1998,defrancoPRB2018} while for $J_H=0$ correlations are AFM and antiferroorbital~\cite{KitaPRB2009}. 

Previous calculations in several lattices, mostly restricted to single-orbital models, have found that including the effect of short-range inter-site correlations in the Mott transition
decreases the critical interactions and can considerably change the expected dependences~\cite{parkPRL2008,MaierRMP2005}. Their effect  in non-ordered states have been studied with cluster approaches~\cite{MaierRMP2005} in the context of organic superconductors and cuprates where they play an important role. Unlike local correlations, these effects do depend on the lattice and become more important when the effective dimension is reduced, as in the honeycomb lattice, with small coordination number $z=3$.

Cluster treatments of multi-orbital models are computationally challenging and very scarce~\cite{PoteryaevPRL2004,BiermannPRL2005,KitaPRB2009,NomuraPRB2014,NomuraPRB2015}. To our knowledge there are no studies which include these non-local correlations in the two-orbital honeycomb lattice.  However, the expected qualitative behaviour in the non-ordered Mott state of the present model can be inferred from previous results in the two-orbital square lattice and in single-orbital models with different lattice geometries.  

As shown in Fig.~\ref{fig:Figtemp}, non-local correlations shift the Mott transition to a smaller critical interaction $U^{c,\rm non-loc}$~\cite{MaierRMP2005,parkPRL2008,schaferPRB2015,liebschPRB2013,KitaPRB2009,LiPRB2015} even in the absence of magnetic order.
The correlated insulator decreases its energy with respect to the correlated metal. 
While for large $T$ and $U$ the local approximation gives a good description of the electronic properties, at lower temperatures and close to $U^{c, \rm non-loc}$ (green line on the left in Fig.~\ref{fig:Figtemp}), the behaviour is controlled by the inter-site correlations.
In the single-orbital case at half-filling and with AFM correlations, the entropy of the insulator decreases due to short-range singlet formation~\cite{parkPRL2008}. 
As a consequence, the critical interaction $U^{c, \rm non-loc}$ reverses its slope as a function of $T$ with respect to the local-correlations prediction~\cite{schaferPRB2015,parkPRL2008,WuPRB2010} (compare blue and green lines in Fig.~\ref{fig:Figtemp}). Hence, the low $T$ phase is the insulator, as observed experimentally. Similar behaviour has been found in a two-orbital model at zero and finite $J_H$~\cite{NomuraPRB2014}.   Cluster DMFT (CDMFT) calculations in the paramagnetic two-orbital square lattice at $x=1/4$ and $J_H=0$~\cite{KitaPRB2009} have also found that the AFM correlations are suppressed by temperature. With increasing $T$ the crossover between the metal and the insulator is non-monotonic, see Fig.~\ref{fig:Figtemp}, and it matches the crossover line found in the single-site approximation at high temperatures (both dashed lines merge). The qualitative behaviour when non-local correlations are included is also found for other lattices. In particular, it is expected to be found in the honeycomb lattice, bipartite as the square one. This has been already studied in the single-orbital case at half-filling~\cite{parkPRL2008,liebschPRB2013}. A larger degree of magnetic frustration decreases the magnitude of the effect, i.e. $U^{c, \rm non-loc}$ becomes closer to the local predictions and the range of temperature in which non-local correlations control the experimental behaviour gets narrower~\cite{TakumaPRL2008}.
We emphasize that the temperature dependence compatible with experimental observations is found only very close to $U^{c, \rm non-loc}$, at interactions for which the system is metallic in the absence of inter-site correlations (namely, below the green line in Fig.~\ref{fig:Figtemp}). This result suggests that inter-site correlations play a key role in the localization and insulating states observed experimentally in TBG. 

The gap size controversy can also be accounted for by the inclusion of non-local correlations. While for local correlations the quasiparticle peak disappears at the Mott transition at $U^{c, \rm local}$ and the gap is the one between the Hubbard bands, for  $U^{c,\rm non-loc}<U<U^{c, \rm local}$ (namely, between the green and blue lines in Fig.~\ref{fig:Figtemp}) non-local correlations open a small gap in the quasiparticle peak for smaller interactions~\cite{parkPRL2008,LiebschPRB2011}. Close to the transition this gap is much smaller than the one between the Hubbard bands and its size would be consistent with experimental observations.

Close to $U^{c,\rm non-loc}$, in the region of parameters where the temperature dependence and gap size place the experimental system, the insulating state is controlled by the inter-site magnetic and orbital correlations. If these correlations are AFM, as it happens for zero or negative Hund~\cite{momoiPRB1998,KitaPRB2009}, a sufficiently large magnetic field could suppress them and induce a transition to a metallic state, as observed experimentally. A magnetic field does not suppress ferromagnetic correlations. As a consequence if the inter-site correlations were ferromagnetic a magnetic field could not produce an insulator to metal transition, in disagreement with experiments. This indicates that the inter-site correlations which assist the Mott transition have antiferromagnetic character. 

Finally, we note that non-local correlations can alter the ratio, discussed above, between the critical interactions at quarter- and half-fillings. The critical interaction is controlled by the non-local correlations which have a different effect at each filling, as it becomes evident by their different ordering tendencies. For example, 
at $J_H=0$ the ground state of the honeycomb lattice is  a valence bond solid~\cite{zhouPRB2016} at $x=1/2$   but a  quantum spin-orbital liquid~\cite{CorbozPRX2012,JakabPRB2016}  at $x=1/4$ in the strong coupling limit. Specific calculations should address whether the inequality 
$U^{c,\rm non-loc}_{1/4}<U^{c,\rm non-loc}_{1/2}$ can be achieved for zero or negative Hund's coupling in the honeycomb lattice.   

\section{Conclusion and Discussion}

We have explored the phenomenology of the insulating states of twisted bilayer graphene 
with $\pm 2$ electrons with respect to the charge neutrality point.
We have shown that the experimental observations dependences are 
not in agreement with the predictions for a non-ordered Mott state if only local correlations are included in the description but could be consistent once non-local correlations are taken into account. In particular,  the insulator to metal transition with a magnetic field or with temperature and the small gap size are in contradiction  with expectations for Mott insulating states driven by local correlations. Even in the absence of magnetic or orbital order, the inclusion of inter-site correlations reduces the critical interaction for the transition, reverses the dependence of the metal-insulator transition with temperature and strongly reduces the gap size close to the transition. A magnetic field can suppress the inter-site correlations, and consequently the insulating behaviour, if the former have antiferromagnetic character, as for $J_H \leq 0$.

The inequality $U^{c,\rm non-loc}_{1/4}<U^{c,\rm non-loc}_{1/2}$, necessary to reproduce the metallic character of the charge neutrality point while the system is insulating at quarter-filling, is fulfilled in the local approximation, but could be affected by non-local effects. Whether it is satisfied in the presence of inter-site correlations should be studied in future work.
An interesting aspect of non-local correlations is the possibility of finding
pseudogap physics. Local correlations result in a momentum-independent self-energy\cite{parkPRL2008}. 
Non-local correlations make the self-energy momentum dependent.  
These momentum dependent correlations have been studied in the single-orbital square lattice when the system is doped away from the insulating state~\cite{ParcolletPRL2004,CivelliPRL2005,KyungPRB2006,TakumaPRL2008,FerreroPRB2009,WernerPRB2009,MerinoPRB2014}. The predicted momentum dependence has been observed in cuprate superconductors and it is a key signature of the pseudogap phenomenology~\cite{ShenScience2005}. 

In this manuscript we have described the TBG with a two-orbital Hubbard model on the honeycomb lattice. The predicted behaviour for the metal insulator transition with temperature and magnetic field, as well as the small gap close to the transition, are not restricted to this model. We expect that our qualitative conclusions would remain valid in other models proposed for the TBG which feature Mott transitions at the experimental fillings.  On the other hand, the range of interaction parameters, temperature and magnetic field in which the inter-site correlations control the experimental behaviour,  the ratio $U^{c,\rm non-loc}_{1/4}/U^{c,\rm non-loc}_{1/2}$, or the possible pseudogap physics will depend on the model. The description of the insulating states with $\pm 1$ and $\pm 3$ electrons from the CNP, recently observed~\cite{yankowitz-arXiv2018}  and not addressed in this work, could require different explanations depending on the lattice model. 
For instance, in a two-orbital honeycomb model these states could correspond to Wigner-Mott insulating states in an extended Hubbard model~\cite{amaricciPRB2010}. 

In summary, our work reconciles the experimental observations for the insulating states with $\pm 2$  electrons with the physics of Mott insulators, in 
agreement with expectations from the exact commensurability. 
The necessity to go beyond the local approximation to reproduce the experiments suggests that  the inclusion of inter-site correlations is not just a more precise way to describe the electronic properties but a key ingredient in the localization of the carriers. The behaviour in a magnetic field indicates that these short-range non-local correlations have antiferromagnetic character.

We thank Pablo San-Jos\'e, Bel\'en Valenzuela, Pablo Jarillo-Herrero, Luca Chirolli, Rafa Rold\'an and Paco Guinea for useful conversations. Funding from Ministerio de Econom\'ia, Industria y Competitividad (Spain) and FEDER funds via grants No. FIS2014-53218-P and FIS2015-64654-P, and from Fundaci\'on Ram\'on Areces is gratefully acknowledged. 

\appendix
\setcounter{section}{0}

\section{Appendix: Critical Interactions} 
\label{app:Uc}
The critical interaction for the Mott transition depends on the charging energy cost $\Delta^U_{x}$, see main text. For the present model, at half-filling, $\Delta^U_{1/2}=U+J_H$ if $J_H \geq 0$ and $\Delta^U_{1/2}= U+5|J_H|$ if $J_H < 0$~\cite{demediciPRB2011,HanPRB1998}. At quarter-filling, $\Delta^U_{1/4}=U-3 J_H$ if $J_H \geq 0$ and $\Delta^U_{1/4}= U$ if $J_H<0$.  
The vanishing density of states at $x=1/2$ pushes the ratio $U^c_{1/2}/U^c_{1/4}\sim1.28$ at $J_H=0$ to slightly larger values than in other lattices~\cite{rozenbergPRB1997,FlorensPRB2004,demediciPRB2011}. For instance, in the square lattice, with a van Hove singularity at $x=1/2$, $U^c_{1/2}/U^c_{1/4} \sim 1.13$. 
Away from $J_H=0$, $U^c_{1/2}$ decreases while $U^c_{1/4}$ has a positive slope, see Fig.~\ref{fig:Fignofield}. These dependences reflect the behaviour of $\Delta^U_x$~\cite{demediciPRB2011}.

\begin{figure}
\leavevmode
\includegraphics[clip,width=0.45\textwidth]{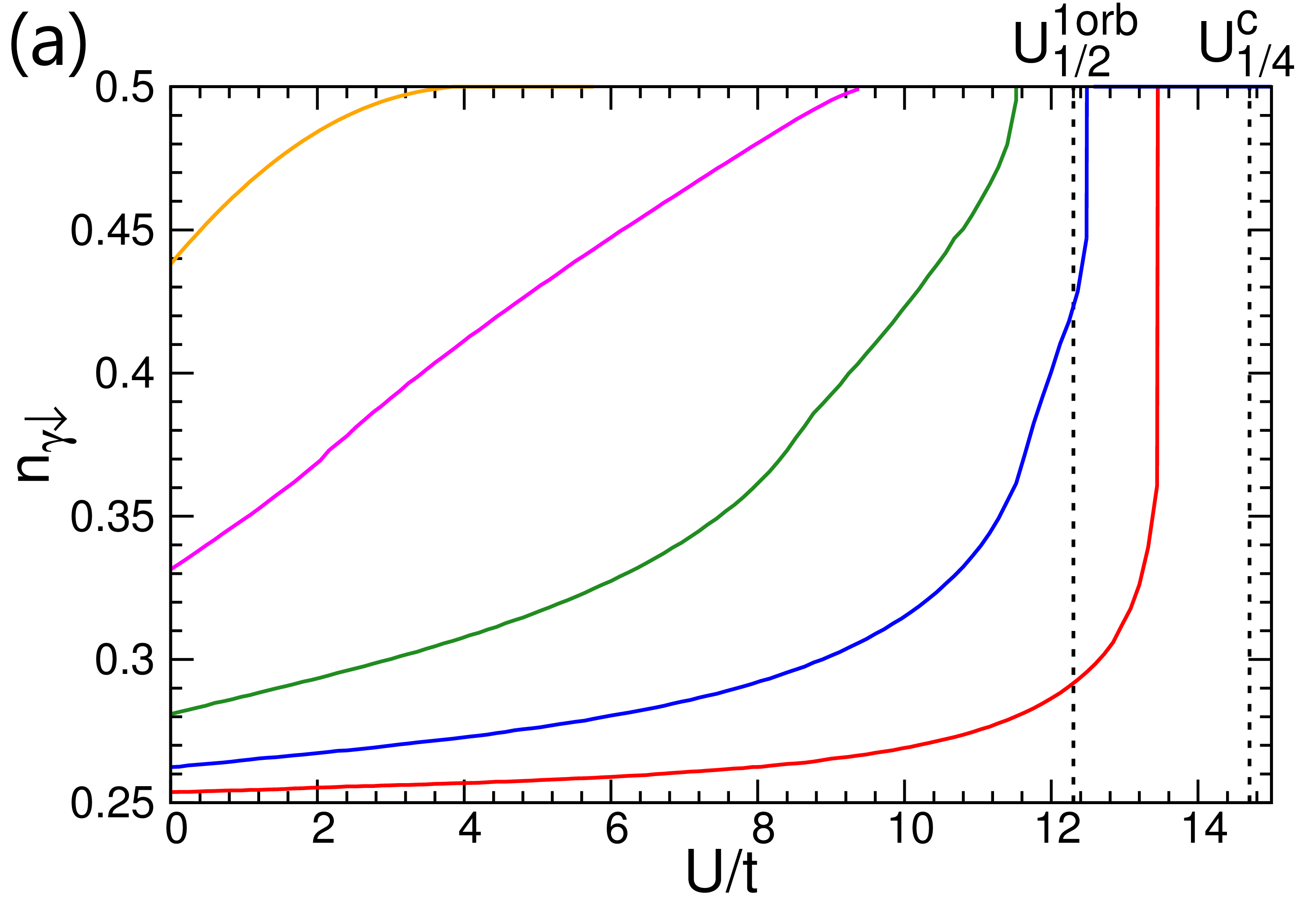}
\includegraphics[clip,width=0.45\textwidth]{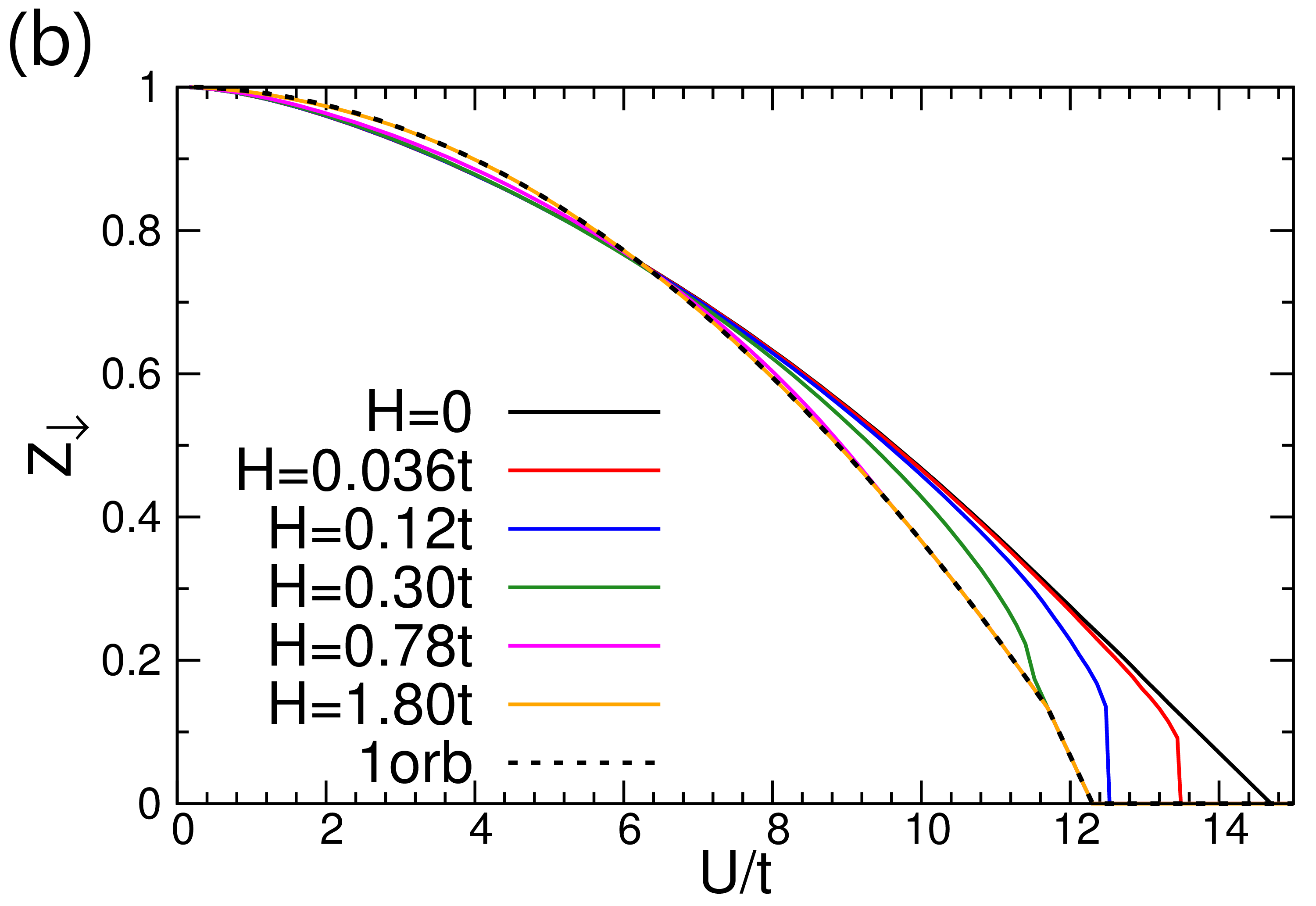}
\caption{(a) Filling per orbital $\gamma$ for the majority (down) spin $n_{\gamma\downarrow}$ as a function of $U$ for $x=1/4$, $J_H=0$, and different values of the magnetic field $H$. $n_{\gamma\uparrow}=0.5-n_{\gamma\downarrow}$. The critical interactions $U^{1\rm orb}_{1/2}$ and $U^c_{1/4}$ are marked with dashed lines. (b) Quasiparticle weight of the majority spin $Z_\downarrow$ for the same values used in (a). 
}  
\label{fig:FigApp} 
\end{figure}

We now focus on $J_H=0$ and $x=1/4$. In the metallic state the strength of the correlations can be quantified by the value of the quasiparticle weight $Z$. $Z=1$ in the non-interacting limit and it decreases with increasing interactions. At the Mott transition $Z$ vanishes. As seen in Fig.~\ref{fig:FigApp}(b), in the absence of a magnetic field, $Z$ vanishes at interactions $U^c_{1/4}$(H=0) larger than $U^{1\rm orb}_{1/2}$.

In the presence of a magnetic field the bands become spin polarized and the quasiparticle weight 
becomes spin dependent $Z_\sigma$. The Mott transition happens when the quasiparticle weight of the majority spin band $Z_\downarrow$ goes to zero.  If the orbitals are completely spin polarized the system becomes equivalent to a single orbital model at $x=1/2$ with interactions $U_{1\rm orb}=U$ for $J_H=0$.  
Except for very small $H$ the spin up band is emptied at interactions $U_{\rm pol} < U^{1\rm orb}_{1/2}$, as it can be seen in Fig.~\ref{fig:FigApp}(a). Beyond $U_{\rm pol}$ the evolution of $Z_\downarrow$ follows the one expected in a single orbital model and $U^c_{1/4}(H)$ tends to  $U^{1\rm orb}_{1/2}$, see Fig.~\ref{fig:FigApp}(b) and Fig.~\ref{fig:Figfield} in the main text. For small fields, $U_{\rm pol}>U^{1\rm orb}_{1/2}$ and $Z_\downarrow$ drops to zero at $U^c_{1/4}(H)=U_{\rm pol}$. 

\bibliography{TBG-tbmott-JPhysComm_arxiv}

\begin{thebibliography}{66}
\expandafter\ifx\csname natexlab\endcsname\relax\def\natexlab#1{#1}\fi
\expandafter\ifx\csname bibnamefont\endcsname\relax
  \def\bibnamefont#1{#1}\fi
\expandafter\ifx\csname bibfnamefont\endcsname\relax
  \def\bibfnamefont#1{#1}\fi
\expandafter\ifx\csname citenamefont\endcsname\relax
  \def\citenamefont#1{#1}\fi
\expandafter\ifx\csname url\endcsname\relax
  \def\url#1{\texttt{#1}}\fi
\expandafter\ifx\csname urlprefix\endcsname\relax\def\urlprefix{URL }\fi
\providecommand{\bibinfo}[2]{#2}
\providecommand{\eprint}[2][]{\url{#2}}

\bibitem[{\citenamefont{Cao et~al.}(2018{\natexlab{a}})\citenamefont{Cao,
  Fatemi, Demir, Fang, Tomarken, Luo, Sanchez-Yamagishi, Watanabe, Taniguchi,
  Kaxiras et~al.}}]{caoNat2018_1}
\bibinfo{author}{\bibfnamefont{Y.}~\bibnamefont{Cao}},
  \bibinfo{author}{\bibfnamefont{V.}~\bibnamefont{Fatemi}},
  \bibinfo{author}{\bibfnamefont{A.}~\bibnamefont{Demir}},
  \bibinfo{author}{\bibfnamefont{S.}~\bibnamefont{Fang}},
  \bibinfo{author}{\bibfnamefont{S.~L.} \bibnamefont{Tomarken}},
  \bibinfo{author}{\bibfnamefont{J.~Y.} \bibnamefont{Luo}},
  \bibinfo{author}{\bibfnamefont{J.~D.} \bibnamefont{Sanchez-Yamagishi}},
  \bibinfo{author}{\bibfnamefont{K.}~\bibnamefont{Watanabe}},
  \bibinfo{author}{\bibfnamefont{T.}~\bibnamefont{Taniguchi}},
  \bibinfo{author}{\bibfnamefont{E.}~\bibnamefont{Kaxiras}},
  \bibnamefont{et~al.}, \bibinfo{journal}{Nature}
  \textbf{\bibinfo{volume}{556}}, \bibinfo{pages}{80}
  (\bibinfo{year}{2018}{\natexlab{a}}).

\bibitem[{\citenamefont{Yankowitz et~al.}(2018)\citenamefont{Yankowitz, Chen,
  Polshyn, Watanabe, Taniguchi, Graf, Young, and Dean}}]{yankowitz-arXiv2018}
\bibinfo{author}{\bibfnamefont{M.}~\bibnamefont{Yankowitz}},
  \bibinfo{author}{\bibfnamefont{S.}~\bibnamefont{Chen}},
  \bibinfo{author}{\bibfnamefont{H.}~\bibnamefont{Polshyn}},
  \bibinfo{author}{\bibfnamefont{K.}~\bibnamefont{Watanabe}},
  \bibinfo{author}{\bibfnamefont{T.}~\bibnamefont{Taniguchi}},
  \bibinfo{author}{\bibfnamefont{D.}~\bibnamefont{Graf}},
  \bibinfo{author}{\bibfnamefont{A.~F.} \bibnamefont{Young}}, \bibnamefont{and}
  \bibinfo{author}{\bibfnamefont{C.~R.} \bibnamefont{Dean}},
  \bibinfo{journal}{arXiv:1808.07865}  (\bibinfo{year}{2018}).

\bibitem[{jar()}]{jarillobeijing}
\bibinfo{note}{P. Jarillo-Herrero, Plenary talk at M2S-2018, Beijing (China).}

\bibitem[{\citenamefont{Cao et~al.}(2018{\natexlab{b}})\citenamefont{Cao,
  Fatemi, Fang, Watanabe, Taniguchi, Kaxiras, and
  Jarillo-Herrero}}]{caoNat2018_2}
\bibinfo{author}{\bibfnamefont{Y.}~\bibnamefont{Cao}},
  \bibinfo{author}{\bibfnamefont{V.}~\bibnamefont{Fatemi}},
  \bibinfo{author}{\bibfnamefont{S.}~\bibnamefont{Fang}},
  \bibinfo{author}{\bibfnamefont{K.}~\bibnamefont{Watanabe}},
  \bibinfo{author}{\bibfnamefont{T.}~\bibnamefont{Taniguchi}},
  \bibinfo{author}{\bibfnamefont{E.}~\bibnamefont{Kaxiras}}, \bibnamefont{and}
  \bibinfo{author}{\bibfnamefont{P.}~\bibnamefont{Jarillo-Herrero}},
  \bibinfo{journal}{Nature} \textbf{\bibinfo{volume}{556}}, \bibinfo{pages}{43}
  (\bibinfo{year}{2018}{\natexlab{b}}).

\bibitem[{\citenamefont{Lopes~dos Santos et~al.}(2007)\citenamefont{Lopes~dos
  Santos, Peres, and Castro~Neto}}]{dossantosPRL2007}
\bibinfo{author}{\bibfnamefont{J.~M.~B.} \bibnamefont{Lopes~dos Santos}},
  \bibinfo{author}{\bibfnamefont{N.~M.~R.} \bibnamefont{Peres}},
  \bibnamefont{and} \bibinfo{author}{\bibfnamefont{A.~H.}
  \bibnamefont{Castro~Neto}}, \bibinfo{journal}{Phys. Rev. Lett.}
  \textbf{\bibinfo{volume}{99}}, \bibinfo{pages}{256802}
  (\bibinfo{year}{2007}).

\bibitem[{\citenamefont{Bistritzer and MacDonald}(2011)}]{bistritzerPNAS2011}
\bibinfo{author}{\bibfnamefont{R.}~\bibnamefont{Bistritzer}} \bibnamefont{and}
  \bibinfo{author}{\bibfnamefont{A.~H.} \bibnamefont{MacDonald}},
  \bibinfo{journal}{Proceedings of the National Academy of Sciences}
  \textbf{\bibinfo{volume}{108}}, \bibinfo{pages}{12233}
  (\bibinfo{year}{2011}).

\bibitem[{\citenamefont{Trambly~de Laissardiere
  et~al.}(2010)\citenamefont{Trambly~de Laissardiere, Mayou, and
  Magaud}}]{LaissardiereNanoLett2010}
\bibinfo{author}{\bibfnamefont{G.}~\bibnamefont{Trambly~de Laissardiere}},
  \bibinfo{author}{\bibfnamefont{D.}~\bibnamefont{Mayou}}, \bibnamefont{and}
  \bibinfo{author}{\bibfnamefont{L.}~\bibnamefont{Magaud}},
  \bibinfo{journal}{Nano Lett} \textbf{\bibinfo{volume}{10}},
  \bibinfo{pages}{804} (\bibinfo{year}{2010}).

\bibitem[{\citenamefont{Yuan and Fu}(2018)}]{yuanPRB2018}
\bibinfo{author}{\bibfnamefont{N.~F.~Q.} \bibnamefont{Yuan}} \bibnamefont{and}
  \bibinfo{author}{\bibfnamefont{L.}~\bibnamefont{Fu}}, \bibinfo{journal}{Phys.
  Rev. B} \textbf{\bibinfo{volume}{98}}, \bibinfo{pages}{045103}
  (\bibinfo{year}{2018}).

\bibitem[{\citenamefont{Zhang}(2018)}]{zhang-arXiv2018}
\bibinfo{author}{\bibfnamefont{L.}~\bibnamefont{Zhang}},
  \bibinfo{journal}{arXiv:1804.09047}  (\bibinfo{year}{2018}).

\bibitem[{\citenamefont{Po et~al.}(2018{\natexlab{a}})\citenamefont{Po, Zou,
  Vishwanath, and Senthil}}]{poPRX2018}
\bibinfo{author}{\bibfnamefont{H.~C.} \bibnamefont{Po}},
  \bibinfo{author}{\bibfnamefont{L.}~\bibnamefont{Zou}},
  \bibinfo{author}{\bibfnamefont{A.}~\bibnamefont{Vishwanath}},
  \bibnamefont{and} \bibinfo{author}{\bibfnamefont{T.}~\bibnamefont{Senthil}},
  \bibinfo{journal}{Phys. Rev. X} \textbf{\bibinfo{volume}{8}},
  \bibinfo{pages}{031089} (\bibinfo{year}{2018}{\natexlab{a}}).

\bibitem[{\citenamefont{Liu et~al.}(2018)\citenamefont{Liu, Zhang, Chen, and
  Yang}}]{liu-arXiv2018}
\bibinfo{author}{\bibfnamefont{C.-C.} \bibnamefont{Liu}},
  \bibinfo{author}{\bibfnamefont{L.-D.} \bibnamefont{Zhang}},
  \bibinfo{author}{\bibfnamefont{W.-Q.} \bibnamefont{Chen}}, \bibnamefont{and}
  \bibinfo{author}{\bibfnamefont{F.}~\bibnamefont{Yang}},
  \bibinfo{journal}{Phys. Rev. Lett.} \textbf{\bibinfo{volume}{121}},
  \bibinfo{pages}{217001} (\bibinfo{year}{2018}).

\bibitem[{\citenamefont{Roy and Juricic}(2018)}]{roy-arXiv2018}
\bibinfo{author}{\bibfnamefont{B.}~\bibnamefont{Roy}} \bibnamefont{and}
  \bibinfo{author}{\bibfnamefont{V.}~\bibnamefont{Juricic}},
  \bibinfo{journal}{arXiv:1803.11190}  (\bibinfo{year}{2018}).

\bibitem[{\citenamefont{Isobe et~al.}(2018)\citenamefont{Isobe, Yuan, and
  Fu}}]{isobe-arXiv2018}
\bibinfo{author}{\bibfnamefont{H.}~\bibnamefont{Isobe}},
  \bibinfo{author}{\bibfnamefont{N.}~\bibnamefont{Yuan}}, \bibnamefont{and}
  \bibinfo{author}{\bibfnamefont{L.}~\bibnamefont{Fu}},
  \bibinfo{journal}{arXiv:1805.06449}  (\bibinfo{year}{2018}).

\bibitem[{\citenamefont{Dodaro et~al.}(2018)\citenamefont{Dodaro, Kivelson,
  Schattner, Sun, and Wang}}]{dodaroPRB2018}
\bibinfo{author}{\bibfnamefont{J.~F.} \bibnamefont{Dodaro}},
  \bibinfo{author}{\bibfnamefont{S.~A.} \bibnamefont{Kivelson}},
  \bibinfo{author}{\bibfnamefont{Y.}~\bibnamefont{Schattner}},
  \bibinfo{author}{\bibfnamefont{X.~Q.} \bibnamefont{Sun}}, \bibnamefont{and}
  \bibinfo{author}{\bibfnamefont{C.}~\bibnamefont{Wang}},
  \bibinfo{journal}{Phys. Rev. B} \textbf{\bibinfo{volume}{98}},
  \bibinfo{pages}{075154} (\bibinfo{year}{2018}).

\bibitem[{\citenamefont{Koshino et~al.}(2018)\citenamefont{Koshino, Yuan,
  Koretsune, Ochi, Kuroki, and Fu}}]{koshinoPRX2018}
\bibinfo{author}{\bibfnamefont{M.}~\bibnamefont{Koshino}},
  \bibinfo{author}{\bibfnamefont{N.~F.~Q.} \bibnamefont{Yuan}},
  \bibinfo{author}{\bibfnamefont{T.}~\bibnamefont{Koretsune}},
  \bibinfo{author}{\bibfnamefont{M.}~\bibnamefont{Ochi}},
  \bibinfo{author}{\bibfnamefont{K.}~\bibnamefont{Kuroki}}, \bibnamefont{and}
  \bibinfo{author}{\bibfnamefont{L.}~\bibnamefont{Fu}}, \bibinfo{journal}{Phys.
  Rev. X} \textbf{\bibinfo{volume}{8}}, \bibinfo{pages}{031087}
  (\bibinfo{year}{2018}).

\bibitem[{\citenamefont{Rozenberg}(1997)}]{rozenbergPRB1997}
\bibinfo{author}{\bibfnamefont{M.~J.} \bibnamefont{Rozenberg}},
  \bibinfo{journal}{Phys. Rev. B} \textbf{\bibinfo{volume}{55}},
  \bibinfo{pages}{R4855} (\bibinfo{year}{1997}).

\bibitem[{\citenamefont{Georges et~al.}(1996)\citenamefont{Georges, Kotliar,
  Krauth, and Rozenberg}}]{georgesRMP1996}
\bibinfo{author}{\bibfnamefont{A.}~\bibnamefont{Georges}},
  \bibinfo{author}{\bibfnamefont{G.}~\bibnamefont{Kotliar}},
  \bibinfo{author}{\bibfnamefont{W.}~\bibnamefont{Krauth}}, \bibnamefont{and}
  \bibinfo{author}{\bibfnamefont{M.~J.} \bibnamefont{Rozenberg}},
  \bibinfo{journal}{Rev. Mod. Phys.} \textbf{\bibinfo{volume}{68}},
  \bibinfo{pages}{13} (\bibinfo{year}{1996}).

\bibitem[{\citenamefont{Kotliar and Ruckenstein}(1986)}]{KotliarPRL1986}
\bibinfo{author}{\bibfnamefont{G.}~\bibnamefont{Kotliar}} \bibnamefont{and}
  \bibinfo{author}{\bibfnamefont{A.~E.} \bibnamefont{Ruckenstein}},
  \bibinfo{journal}{Phys. Rev. Lett.} \textbf{\bibinfo{volume}{57}},
  \bibinfo{pages}{1362} (\bibinfo{year}{1986}).

\bibitem[{\citenamefont{Florens and Georges}(2004)}]{FlorensPRB2004}
\bibinfo{author}{\bibfnamefont{S.}~\bibnamefont{Florens}} \bibnamefont{and}
  \bibinfo{author}{\bibfnamefont{A.}~\bibnamefont{Georges}},
  \bibinfo{journal}{Phys. Rev. B} \textbf{\bibinfo{volume}{70}},
  \bibinfo{pages}{035114} (\bibinfo{year}{2004}).

\bibitem[{\citenamefont{{\mbox de' Medici} et~al.}(2005)\citenamefont{{\mbox
  de' Medici}, Georges, and Biermann}}]{demediciPRB2005}
\bibinfo{author}{\bibfnamefont{L.}~\bibnamefont{{\mbox de' Medici}}},
  \bibinfo{author}{\bibfnamefont{A.}~\bibnamefont{Georges}}, \bibnamefont{and}
  \bibinfo{author}{\bibfnamefont{S.}~\bibnamefont{Biermann}},
  \bibinfo{journal}{Phys. Rev. B} \textbf{\bibinfo{volume}{72}},
  \bibinfo{pages}{205124} (\bibinfo{year}{2005}).

\bibitem[{\citenamefont{Yu and Si}(2012)}]{yuPRB2012}
\bibinfo{author}{\bibfnamefont{R.}~\bibnamefont{Yu}} \bibnamefont{and}
  \bibinfo{author}{\bibfnamefont{Q.}~\bibnamefont{Si}}, \bibinfo{journal}{Phys.
  Rev. B} \textbf{\bibinfo{volume}{86}}, \bibinfo{pages}{085104}
  (\bibinfo{year}{2012}).

\bibitem[{\citenamefont{Kang and Vafek}(2018)}]{kangPRX2018}
\bibinfo{author}{\bibfnamefont{J.}~\bibnamefont{Kang}} \bibnamefont{and}
  \bibinfo{author}{\bibfnamefont{O.}~\bibnamefont{Vafek}},
  \bibinfo{journal}{Phys. Rev. X} \textbf{\bibinfo{volume}{8}},
  \bibinfo{pages}{031088} (\bibinfo{year}{2018}).

\bibitem[{\citenamefont{Rademaker and Mellado}(2018)}]{rademaker-arXiv2018}
\bibinfo{author}{\bibfnamefont{L.}~\bibnamefont{Rademaker}} \bibnamefont{and}
  \bibinfo{author}{\bibfnamefont{P.}~\bibnamefont{Mellado}},
  \bibinfo{journal}{arXiv:1805.05294}  (\bibinfo{year}{2018}).

\bibitem[{\citenamefont{Zou et~al.}(2018)\citenamefont{Zou, Po, Vishwanath, and
  Senthil}}]{ZouPRB2018}
\bibinfo{author}{\bibfnamefont{L.}~\bibnamefont{Zou}},
  \bibinfo{author}{\bibfnamefont{H.~C.} \bibnamefont{Po}},
  \bibinfo{author}{\bibfnamefont{A.}~\bibnamefont{Vishwanath}},
  \bibnamefont{and} \bibinfo{author}{\bibfnamefont{T.}~\bibnamefont{Senthil}},
  \bibinfo{journal}{Phys. Rev. B} \textbf{\bibinfo{volume}{98}},
  \bibinfo{pages}{085435} (\bibinfo{year}{2018}).

\bibitem[{\citenamefont{Po et~al.}(2018{\natexlab{b}})\citenamefont{Po, Zou,
  Senthil, and Vishwanath}}]{Po-arXiv2018-2}
\bibinfo{author}{\bibfnamefont{H.~C.} \bibnamefont{Po}},
  \bibinfo{author}{\bibfnamefont{L.}~\bibnamefont{Zou}},
  \bibinfo{author}{\bibfnamefont{T.}~\bibnamefont{Senthil}}, \bibnamefont{and}
  \bibinfo{author}{\bibfnamefont{A.}~\bibnamefont{Vishwanath}},
  \bibinfo{journal}{arXiv:1808.02482}  (\bibinfo{year}{2018}{\natexlab{b}}).

\bibitem[{\citenamefont{Angeli et~al.}(2018)\citenamefont{Angeli, Mandelli,
  Valli, Amaricci, Capone, Tosatti, and Fabrizio}}]{Angeli-arXiv2018}
\bibinfo{author}{\bibfnamefont{M.}~\bibnamefont{Angeli}},
  \bibinfo{author}{\bibfnamefont{D.}~\bibnamefont{Mandelli}},
  \bibinfo{author}{\bibfnamefont{A.}~\bibnamefont{Valli}},
  \bibinfo{author}{\bibfnamefont{A.}~\bibnamefont{Amaricci}},
  \bibinfo{author}{\bibfnamefont{M.}~\bibnamefont{Capone}},
  \bibinfo{author}{\bibfnamefont{E.}~\bibnamefont{Tosatti}}, \bibnamefont{and}
  \bibinfo{author}{\bibfnamefont{M.}~\bibnamefont{Fabrizio}},
  \bibinfo{journal}{arXiv:1809.11140}  (\bibinfo{year}{2018}).

\bibitem[{\citenamefont{Castellani et~al.}(1978)\citenamefont{Castellani,
  Natoli, and Ranninger}}]{castellaniPRB1978}
\bibinfo{author}{\bibfnamefont{C.}~\bibnamefont{Castellani}},
  \bibinfo{author}{\bibfnamefont{C.~R.} \bibnamefont{Natoli}},
  \bibnamefont{and}
  \bibinfo{author}{\bibfnamefont{J.}~\bibnamefont{Ranninger}},
  \bibinfo{journal}{Phys. Rev. B} \textbf{\bibinfo{volume}{18}},
  \bibinfo{pages}{4945} (\bibinfo{year}{1978}).

\bibitem[{\citenamefont{Nomura et~al.}(2016)\citenamefont{Nomura, Sakai,
  Capone, and Arita}}]{nomuraJPCM2016}
\bibinfo{author}{\bibfnamefont{Y.}~\bibnamefont{Nomura}},
  \bibinfo{author}{\bibfnamefont{S.}~\bibnamefont{Sakai}},
  \bibinfo{author}{\bibfnamefont{M.}~\bibnamefont{Capone}}, \bibnamefont{and}
  \bibinfo{author}{\bibfnamefont{R.}~\bibnamefont{Arita}},
  \bibinfo{journal}{Journal of Physics: Condensed Matter}
  \textbf{\bibinfo{volume}{28}}, \bibinfo{pages}{153001}
  (\bibinfo{year}{2016}).

\bibitem[{\citenamefont{de' Medici}(2011)}]{demediciPRB2011}
\bibinfo{author}{\bibfnamefont{L.}~\bibnamefont{de' Medici}},
  \bibinfo{journal}{Phys. Rev. B} \textbf{\bibinfo{volume}{83}},
  \bibinfo{pages}{205112} (\bibinfo{year}{2011}).

\bibitem[{\citenamefont{Yi et~al.}(2013)\citenamefont{Yi, Lu, Yu, Riggs, Chu,
  Lv, Liu, Lu, Cui, Hashimoto et~al.}}]{yiPRL2013}
\bibinfo{author}{\bibfnamefont{M.}~\bibnamefont{Yi}},
  \bibinfo{author}{\bibfnamefont{D.~H.} \bibnamefont{Lu}},
  \bibinfo{author}{\bibfnamefont{R.}~\bibnamefont{Yu}},
  \bibinfo{author}{\bibfnamefont{S.~C.} \bibnamefont{Riggs}},
  \bibinfo{author}{\bibfnamefont{J.-H.} \bibnamefont{Chu}},
  \bibinfo{author}{\bibfnamefont{B.}~\bibnamefont{Lv}},
  \bibinfo{author}{\bibfnamefont{Z.~K.} \bibnamefont{Liu}},
  \bibinfo{author}{\bibfnamefont{M.}~\bibnamefont{Lu}},
  \bibinfo{author}{\bibfnamefont{Y.-T.} \bibnamefont{Cui}},
  \bibinfo{author}{\bibfnamefont{M.}~\bibnamefont{Hashimoto}},
  \bibnamefont{et~al.}, \bibinfo{journal}{Phys. Rev. Lett.}
  \textbf{\bibinfo{volume}{110}}, \bibinfo{pages}{067003}
  (\bibinfo{year}{2013}).

\bibitem[{\citenamefont{de' Medici et~al.}(2014)\citenamefont{de' Medici,
  Giovannetti, and Capone}}]{demediciPRL2014}
\bibinfo{author}{\bibfnamefont{L.}~\bibnamefont{de' Medici}},
  \bibinfo{author}{\bibfnamefont{G.}~\bibnamefont{Giovannetti}},
  \bibnamefont{and} \bibinfo{author}{\bibfnamefont{M.}~\bibnamefont{Capone}},
  \bibinfo{journal}{Phys. Rev. Lett.} \textbf{\bibinfo{volume}{112}},
  \bibinfo{pages}{177001} (\bibinfo{year}{2014}).

\bibitem[{\citenamefont{Calder\'on et~al.}(2014)\citenamefont{Calder\'on, de'
  Medici, Valenzuela, and Bascones}}]{CalderonPRB2014}
\bibinfo{author}{\bibfnamefont{M.~J.} \bibnamefont{Calder\'on}},
  \bibinfo{author}{\bibfnamefont{L.}~\bibnamefont{de' Medici}},
  \bibinfo{author}{\bibfnamefont{B.}~\bibnamefont{Valenzuela}},
  \bibnamefont{and} \bibinfo{author}{\bibfnamefont{E.}~\bibnamefont{Bascones}},
  \bibinfo{journal}{Phys. Rev. B} \textbf{\bibinfo{volume}{90}},
  \bibinfo{pages}{115128} (\bibinfo{year}{2014}).

\bibitem[{\citenamefont{Fanfarillo and Bascones}(2015)}]{FanfarilloPRB2015}
\bibinfo{author}{\bibfnamefont{L.}~\bibnamefont{Fanfarillo}} \bibnamefont{and}
  \bibinfo{author}{\bibfnamefont{E.}~\bibnamefont{Bascones}},
  \bibinfo{journal}{Phys. Rev. B} \textbf{\bibinfo{volume}{92}},
  \bibinfo{pages}{075136} (\bibinfo{year}{2015}).

\bibitem[{\citenamefont{Fanfarillo et~al.}(2017)\citenamefont{Fanfarillo,
  Giovannetti, Capone, and Bascones}}]{FanfarilloPRB2017}
\bibinfo{author}{\bibfnamefont{L.}~\bibnamefont{Fanfarillo}},
  \bibinfo{author}{\bibfnamefont{G.}~\bibnamefont{Giovannetti}},
  \bibinfo{author}{\bibfnamefont{M.}~\bibnamefont{Capone}}, \bibnamefont{and}
  \bibinfo{author}{\bibfnamefont{E.}~\bibnamefont{Bascones}},
  \bibinfo{journal}{Phys. Rev. B} \textbf{\bibinfo{volume}{95}},
  \bibinfo{pages}{144511} (\bibinfo{year}{2017}).

\bibitem[{foo({\natexlab{a}})}]{footnote-DMFT-slavespin}
\bibinfo{note}{The dependence of the critical interaction on the Hund's
  coupling, the filling and the tight-binding model is very well captures by
  slave-spin. Nevertheless slave-spin gives values of $U^c$ slightly larger
  than DMFT. For example, the Mott transition in the single-orbital honeycomb
  lattice is found at $U^c \sim10 t$ in DMFT and at $U^c \sim 12 t$ using
  slave-spin.}

\bibitem[{\citenamefont{Han et~al.}(1998)\citenamefont{Han, Jarrell, and
  Cox}}]{HanPRB1998}
\bibinfo{author}{\bibfnamefont{J.~E.} \bibnamefont{Han}},
  \bibinfo{author}{\bibfnamefont{M.}~\bibnamefont{Jarrell}}, \bibnamefont{and}
  \bibinfo{author}{\bibfnamefont{D.~L.} \bibnamefont{Cox}},
  \bibinfo{journal}{Phys. Rev. B} \textbf{\bibinfo{volume}{58}},
  \bibinfo{pages}{R4199} (\bibinfo{year}{1998}).

\bibitem[{\citenamefont{Vu\ifmmode \check{c}\else \v{c}\fi{}i\ifmmode
  \check{c}\else \v{c}\fi{}evi\ifmmode~\acute{c}\else \'{c}\fi{}
  et~al.}(2013)\citenamefont{Vu\ifmmode \check{c}\else \v{c}\fi{}i\ifmmode
  \check{c}\else \v{c}\fi{}evi\ifmmode~\acute{c}\else \'{c}\fi{}, Terletska,
  Tanaskovi\ifmmode~\acute{c}\else \'{c}\fi{}, and
  Dobrosavljevi\ifmmode~\acute{c}\else \'{c}\fi{}}}]{TerletskaPRB2013}
\bibinfo{author}{\bibfnamefont{J.}~\bibnamefont{Vu\ifmmode \check{c}\else
  \v{c}\fi{}i\ifmmode \check{c}\else \v{c}\fi{}evi\ifmmode~\acute{c}\else
  \'{c}\fi{}}}, \bibinfo{author}{\bibfnamefont{H.}~\bibnamefont{Terletska}},
  \bibinfo{author}{\bibfnamefont{D.}~\bibnamefont{Tanaskovi\ifmmode~\acute{c}\else
  \'{c}\fi{}}}, \bibnamefont{and}
  \bibinfo{author}{\bibfnamefont{V.}~\bibnamefont{Dobrosavljevi\ifmmode~\acute{c}\else
  \'{c}\fi{}}}, \bibinfo{journal}{Phys. Rev. B} \textbf{\bibinfo{volume}{88}},
  \bibinfo{pages}{075143} (\bibinfo{year}{2013}).

\bibitem[{\citenamefont{Fazekas}(1999)}]{fazekas-book}
\bibinfo{author}{\bibfnamefont{P.}~\bibnamefont{Fazekas}},
  \emph{\bibinfo{title}{Lecture notes on Electron Correlations and Magnetism}}
  (\bibinfo{publisher}{World Scientific Publishing Company},
  \bibinfo{year}{1999}).

\bibitem[{\citenamefont{Limelette et~al.}(2003)\citenamefont{Limelette,
  Georges, J{\'e}rome, Wzietek, Metcalf, and Honig}}]{LimeletteScience2003}
\bibinfo{author}{\bibfnamefont{P.}~\bibnamefont{Limelette}},
  \bibinfo{author}{\bibfnamefont{A.}~\bibnamefont{Georges}},
  \bibinfo{author}{\bibfnamefont{D.}~\bibnamefont{J{\'e}rome}},
  \bibinfo{author}{\bibfnamefont{P.}~\bibnamefont{Wzietek}},
  \bibinfo{author}{\bibfnamefont{P.}~\bibnamefont{Metcalf}}, \bibnamefont{and}
  \bibinfo{author}{\bibfnamefont{J.~M.} \bibnamefont{Honig}},
  \bibinfo{journal}{Science} \textbf{\bibinfo{volume}{302}},
  \bibinfo{pages}{89} (\bibinfo{year}{2003}).

\bibitem[{\citenamefont{Park et~al.}(2008)\citenamefont{Park, Haule, and
  Kotliar}}]{parkPRL2008}
\bibinfo{author}{\bibfnamefont{H.}~\bibnamefont{Park}},
  \bibinfo{author}{\bibfnamefont{K.}~\bibnamefont{Haule}}, \bibnamefont{and}
  \bibinfo{author}{\bibfnamefont{G.}~\bibnamefont{Kotliar}},
  \bibinfo{journal}{Phys. Rev. Lett.} \textbf{\bibinfo{volume}{101}},
  \bibinfo{pages}{186403} (\bibinfo{year}{2008}).

\bibitem[{foo({\natexlab{b}})}]{footnote-field1orb}
\bibinfo{note}{In fact, in the local approximation, even in the single orbital
  Hubbard model, a finite magnetic field enhances the insulating
  behavior~\cite{LalouxPRB1994}.}

\bibitem[{\citenamefont{Nomura et~al.}(2014)\citenamefont{Nomura, Sakai, and
  Arita}}]{NomuraPRB2014}
\bibinfo{author}{\bibfnamefont{Y.}~\bibnamefont{Nomura}},
  \bibinfo{author}{\bibfnamefont{S.}~\bibnamefont{Sakai}}, \bibnamefont{and}
  \bibinfo{author}{\bibfnamefont{R.}~\bibnamefont{Arita}},
  \bibinfo{journal}{Phys. Rev. B} \textbf{\bibinfo{volume}{89}},
  \bibinfo{pages}{195146} (\bibinfo{year}{2014}).

\bibitem[{\citenamefont{Nomura et~al.}(2015)\citenamefont{Nomura, Sakai, and
  Arita}}]{NomuraPRB2015}
\bibinfo{author}{\bibfnamefont{Y.}~\bibnamefont{Nomura}},
  \bibinfo{author}{\bibfnamefont{S.}~\bibnamefont{Sakai}}, \bibnamefont{and}
  \bibinfo{author}{\bibfnamefont{R.}~\bibnamefont{Arita}},
  \bibinfo{journal}{Phys. Rev. B} \textbf{\bibinfo{volume}{91}},
  \bibinfo{pages}{235107} (\bibinfo{year}{2015}).

\bibitem[{\citenamefont{Momoi and Kubo}(1998)}]{momoiPRB1998}
\bibinfo{author}{\bibfnamefont{T.}~\bibnamefont{Momoi}} \bibnamefont{and}
  \bibinfo{author}{\bibfnamefont{K.}~\bibnamefont{Kubo}},
  \bibinfo{journal}{Phys. Rev. B} \textbf{\bibinfo{volume}{58}},
  \bibinfo{pages}{R567} (\bibinfo{year}{1998}).

\bibitem[{\citenamefont{De~Franco et~al.}(2018)\citenamefont{De~Franco,
  Tocchio, and Becca}}]{defrancoPRB2018}
\bibinfo{author}{\bibfnamefont{C.}~\bibnamefont{De~Franco}},
  \bibinfo{author}{\bibfnamefont{L.~F.} \bibnamefont{Tocchio}},
  \bibnamefont{and} \bibinfo{author}{\bibfnamefont{F.}~\bibnamefont{Becca}},
  \bibinfo{journal}{Phys. Rev. B} \textbf{\bibinfo{volume}{98}},
  \bibinfo{pages}{075117} (\bibinfo{year}{2018}).

\bibitem[{\citenamefont{Kita et~al.}(2009)\citenamefont{Kita, Ohashi, and
  Suga}}]{KitaPRB2009}
\bibinfo{author}{\bibfnamefont{T.}~\bibnamefont{Kita}},
  \bibinfo{author}{\bibfnamefont{T.}~\bibnamefont{Ohashi}}, \bibnamefont{and}
  \bibinfo{author}{\bibfnamefont{S.-i.} \bibnamefont{Suga}},
  \bibinfo{journal}{Phys. Rev. B} \textbf{\bibinfo{volume}{79}},
  \bibinfo{pages}{245128} (\bibinfo{year}{2009}).

\bibitem[{\citenamefont{Maier et~al.}(2005)\citenamefont{Maier, Jarrell,
  Pruschke, and Hettler}}]{MaierRMP2005}
\bibinfo{author}{\bibfnamefont{T.}~\bibnamefont{Maier}},
  \bibinfo{author}{\bibfnamefont{M.}~\bibnamefont{Jarrell}},
  \bibinfo{author}{\bibfnamefont{T.}~\bibnamefont{Pruschke}}, \bibnamefont{and}
  \bibinfo{author}{\bibfnamefont{M.~H.} \bibnamefont{Hettler}},
  \bibinfo{journal}{Rev. Mod. Phys.} \textbf{\bibinfo{volume}{77}},
  \bibinfo{pages}{1027} (\bibinfo{year}{2005}).

\bibitem[{\citenamefont{Poteryaev et~al.}(2004)\citenamefont{Poteryaev,
  Lichtenstein, and Kotliar}}]{PoteryaevPRL2004}
\bibinfo{author}{\bibfnamefont{A.~I.} \bibnamefont{Poteryaev}},
  \bibinfo{author}{\bibfnamefont{A.~I.} \bibnamefont{Lichtenstein}},
  \bibnamefont{and} \bibinfo{author}{\bibfnamefont{G.}~\bibnamefont{Kotliar}},
  \bibinfo{journal}{Phys. Rev. Lett.} \textbf{\bibinfo{volume}{93}},
  \bibinfo{pages}{086401} (\bibinfo{year}{2004}).

\bibitem[{\citenamefont{Biermann et~al.}(2005)\citenamefont{Biermann,
  Poteryaev, Lichtenstein, and Georges}}]{BiermannPRL2005}
\bibinfo{author}{\bibfnamefont{S.}~\bibnamefont{Biermann}},
  \bibinfo{author}{\bibfnamefont{A.}~\bibnamefont{Poteryaev}},
  \bibinfo{author}{\bibfnamefont{A.~I.} \bibnamefont{Lichtenstein}},
  \bibnamefont{and} \bibinfo{author}{\bibfnamefont{A.}~\bibnamefont{Georges}},
  \bibinfo{journal}{Phys. Rev. Lett.} \textbf{\bibinfo{volume}{94}},
  \bibinfo{pages}{026404} (\bibinfo{year}{2005}).

\bibitem[{\citenamefont{Sch\"afer et~al.}(2015)\citenamefont{Sch\"afer, Geles,
  Rost, Rohringer, Arrigoni, Held, Bl\"umer, Aichhorfn, and
  Toschi}}]{schaferPRB2015}
\bibinfo{author}{\bibfnamefont{T.}~\bibnamefont{Sch\"afer}},
  \bibinfo{author}{\bibfnamefont{F.}~\bibnamefont{Geles}},
  \bibinfo{author}{\bibfnamefont{D.}~\bibnamefont{Rost}},
  \bibinfo{author}{\bibfnamefont{G.}~\bibnamefont{Rohringer}},
  \bibinfo{author}{\bibfnamefont{E.}~\bibnamefont{Arrigoni}},
  \bibinfo{author}{\bibfnamefont{K.}~\bibnamefont{Held}},
  \bibinfo{author}{\bibfnamefont{N.}~\bibnamefont{Bl\"umer}},
  \bibinfo{author}{\bibfnamefont{M.}~\bibnamefont{Aichhorfn}},
  \bibnamefont{and} \bibinfo{author}{\bibfnamefont{A.}~\bibnamefont{Toschi}},
  \bibinfo{journal}{Phys. Rev. B} \textbf{\bibinfo{volume}{91}},
  \bibinfo{pages}{125109} (\bibinfo{year}{2015}).

\bibitem[{\citenamefont{Liebsch and Wu}(2013)}]{liebschPRB2013}
\bibinfo{author}{\bibfnamefont{A.}~\bibnamefont{Liebsch}} \bibnamefont{and}
  \bibinfo{author}{\bibfnamefont{W.}~\bibnamefont{Wu}}, \bibinfo{journal}{Phys.
  Rev. B} \textbf{\bibinfo{volume}{87}}, \bibinfo{pages}{205127}
  (\bibinfo{year}{2013}).

\bibitem[{\citenamefont{Li et~al.}(2015)\citenamefont{Li, He, and
  Lu}}]{LiPRB2015}
\bibinfo{author}{\bibfnamefont{Q.-X.} \bibnamefont{Li}},
  \bibinfo{author}{\bibfnamefont{R.-Q.} \bibnamefont{He}}, \bibnamefont{and}
  \bibinfo{author}{\bibfnamefont{Z.-Y.} \bibnamefont{Lu}},
  \bibinfo{journal}{Phys. Rev. B} \textbf{\bibinfo{volume}{92}},
  \bibinfo{pages}{155127} (\bibinfo{year}{2015}).

\bibitem[{\citenamefont{Wu et~al.}(2010)\citenamefont{Wu, Chen, Tao, Tong, and
  Liu}}]{WuPRB2010}
\bibinfo{author}{\bibfnamefont{W.}~\bibnamefont{Wu}},
  \bibinfo{author}{\bibfnamefont{Y.-H.} \bibnamefont{Chen}},
  \bibinfo{author}{\bibfnamefont{H.-S.} \bibnamefont{Tao}},
  \bibinfo{author}{\bibfnamefont{N.-H.} \bibnamefont{Tong}}, \bibnamefont{and}
  \bibinfo{author}{\bibfnamefont{W.-M.} \bibnamefont{Liu}},
  \bibinfo{journal}{Phys. Rev. B} \textbf{\bibinfo{volume}{82}},
  \bibinfo{pages}{245102} (\bibinfo{year}{2010}).

\bibitem[{\citenamefont{Ohashi et~al.}(2008)\citenamefont{Ohashi, Momoi,
  Tsunetsugu, and Kawakami}}]{TakumaPRL2008}
\bibinfo{author}{\bibfnamefont{T.}~\bibnamefont{Ohashi}},
  \bibinfo{author}{\bibfnamefont{T.}~\bibnamefont{Momoi}},
  \bibinfo{author}{\bibfnamefont{H.}~\bibnamefont{Tsunetsugu}},
  \bibnamefont{and} \bibinfo{author}{\bibfnamefont{N.}~\bibnamefont{Kawakami}},
  \bibinfo{journal}{Phys. Rev. Lett.} \textbf{\bibinfo{volume}{100}},
  \bibinfo{pages}{076402} (\bibinfo{year}{2008}).

\bibitem[{\citenamefont{Liebsch}(2011)}]{LiebschPRB2011}
\bibinfo{author}{\bibfnamefont{A.}~\bibnamefont{Liebsch}},
  \bibinfo{journal}{Phys. Rev. B} \textbf{\bibinfo{volume}{83}},
  \bibinfo{pages}{035113} (\bibinfo{year}{2011}).

\bibitem[{\citenamefont{Zhou et~al.}(2016)\citenamefont{Zhou, Wang, Meng, Wang,
  and Wu}}]{zhouPRB2016}
\bibinfo{author}{\bibfnamefont{Z.}~\bibnamefont{Zhou}},
  \bibinfo{author}{\bibfnamefont{D.}~\bibnamefont{Wang}},
  \bibinfo{author}{\bibfnamefont{Z.~Y.} \bibnamefont{Meng}},
  \bibinfo{author}{\bibfnamefont{Y.}~\bibnamefont{Wang}}, \bibnamefont{and}
  \bibinfo{author}{\bibfnamefont{C.}~\bibnamefont{Wu}}, \bibinfo{journal}{Phys.
  Rev. B} \textbf{\bibinfo{volume}{93}}, \bibinfo{pages}{245157}
  (\bibinfo{year}{2016}).

\bibitem[{\citenamefont{Corboz et~al.}(2012)\citenamefont{Corboz, Lajk\'o,
  L\"auchli, Penc, and Mila}}]{CorbozPRX2012}
\bibinfo{author}{\bibfnamefont{P.}~\bibnamefont{Corboz}},
  \bibinfo{author}{\bibfnamefont{M.}~\bibnamefont{Lajk\'o}},
  \bibinfo{author}{\bibfnamefont{A.~M.} \bibnamefont{L\"auchli}},
  \bibinfo{author}{\bibfnamefont{K.}~\bibnamefont{Penc}}, \bibnamefont{and}
  \bibinfo{author}{\bibfnamefont{F.}~\bibnamefont{Mila}},
  \bibinfo{journal}{Phys. Rev. X} \textbf{\bibinfo{volume}{2}},
  \bibinfo{pages}{041013} (\bibinfo{year}{2012}).

\bibitem[{\citenamefont{Jakab et~al.}(2016)\citenamefont{Jakab, Szirmai,
  Lewenstein, and Szirmai}}]{JakabPRB2016}
\bibinfo{author}{\bibfnamefont{D.}~\bibnamefont{Jakab}},
  \bibinfo{author}{\bibfnamefont{E.}~\bibnamefont{Szirmai}},
  \bibinfo{author}{\bibfnamefont{M.}~\bibnamefont{Lewenstein}},
  \bibnamefont{and} \bibinfo{author}{\bibfnamefont{G.}~\bibnamefont{Szirmai}},
  \bibinfo{journal}{Phys. Rev. B} \textbf{\bibinfo{volume}{93}},
  \bibinfo{pages}{064434} (\bibinfo{year}{2016}).

\bibitem[{\citenamefont{Parcollet et~al.}(2004)\citenamefont{Parcollet, Biroli,
  and Kotliar}}]{ParcolletPRL2004}
\bibinfo{author}{\bibfnamefont{O.}~\bibnamefont{Parcollet}},
  \bibinfo{author}{\bibfnamefont{G.}~\bibnamefont{Biroli}}, \bibnamefont{and}
  \bibinfo{author}{\bibfnamefont{G.}~\bibnamefont{Kotliar}},
  \bibinfo{journal}{Phys. Rev. Lett.} \textbf{\bibinfo{volume}{92}},
  \bibinfo{pages}{226402} (\bibinfo{year}{2004}).

\bibitem[{\citenamefont{Civelli et~al.}(2005)\citenamefont{Civelli, Capone,
  Kancharla, Parcollet, and Kotliar}}]{CivelliPRL2005}
\bibinfo{author}{\bibfnamefont{M.}~\bibnamefont{Civelli}},
  \bibinfo{author}{\bibfnamefont{M.}~\bibnamefont{Capone}},
  \bibinfo{author}{\bibfnamefont{S.~S.} \bibnamefont{Kancharla}},
  \bibinfo{author}{\bibfnamefont{O.}~\bibnamefont{Parcollet}},
  \bibnamefont{and} \bibinfo{author}{\bibfnamefont{G.}~\bibnamefont{Kotliar}},
  \bibinfo{journal}{Phys. Rev. Lett.} \textbf{\bibinfo{volume}{95}},
  \bibinfo{pages}{106402} (\bibinfo{year}{2005}).

\bibitem[{\citenamefont{Kyung et~al.}(2006)\citenamefont{Kyung, Kancharla,
  S\'en\'echal, Tremblay, Civelli, and Kotliar}}]{KyungPRB2006}
\bibinfo{author}{\bibfnamefont{B.}~\bibnamefont{Kyung}},
  \bibinfo{author}{\bibfnamefont{S.~S.} \bibnamefont{Kancharla}},
  \bibinfo{author}{\bibfnamefont{D.}~\bibnamefont{S\'en\'echal}},
  \bibinfo{author}{\bibfnamefont{A.-M.~S.} \bibnamefont{Tremblay}},
  \bibinfo{author}{\bibfnamefont{M.}~\bibnamefont{Civelli}}, \bibnamefont{and}
  \bibinfo{author}{\bibfnamefont{G.}~\bibnamefont{Kotliar}},
  \bibinfo{journal}{Phys. Rev. B} \textbf{\bibinfo{volume}{73}},
  \bibinfo{pages}{165114} (\bibinfo{year}{2006}).

\bibitem[{\citenamefont{Ferrero et~al.}(2009)\citenamefont{Ferrero, Cornaglia,
  De~Leo, Parcollet, Kotliar, and Georges}}]{FerreroPRB2009}
\bibinfo{author}{\bibfnamefont{M.}~\bibnamefont{Ferrero}},
  \bibinfo{author}{\bibfnamefont{P.~S.} \bibnamefont{Cornaglia}},
  \bibinfo{author}{\bibfnamefont{L.}~\bibnamefont{De~Leo}},
  \bibinfo{author}{\bibfnamefont{O.}~\bibnamefont{Parcollet}},
  \bibinfo{author}{\bibfnamefont{G.}~\bibnamefont{Kotliar}}, \bibnamefont{and}
  \bibinfo{author}{\bibfnamefont{A.}~\bibnamefont{Georges}},
  \bibinfo{journal}{Phys. Rev. B} \textbf{\bibinfo{volume}{80}},
  \bibinfo{pages}{064501} (\bibinfo{year}{2009}).

\bibitem[{\citenamefont{Werner et~al.}(2009)\citenamefont{Werner, Gull,
  Parcollet, and Millis}}]{WernerPRB2009}
\bibinfo{author}{\bibfnamefont{P.}~\bibnamefont{Werner}},
  \bibinfo{author}{\bibfnamefont{E.}~\bibnamefont{Gull}},
  \bibinfo{author}{\bibfnamefont{O.}~\bibnamefont{Parcollet}},
  \bibnamefont{and} \bibinfo{author}{\bibfnamefont{A.~J.}
  \bibnamefont{Millis}}, \bibinfo{journal}{Phys. Rev. B}
  \textbf{\bibinfo{volume}{80}}, \bibinfo{pages}{045120}
  (\bibinfo{year}{2009}).

\bibitem[{\citenamefont{Merino and Gunnarsson}(2014)}]{MerinoPRB2014}
\bibinfo{author}{\bibfnamefont{J.}~\bibnamefont{Merino}} \bibnamefont{and}
  \bibinfo{author}{\bibfnamefont{O.}~\bibnamefont{Gunnarsson}},
  \bibinfo{journal}{Phys. Rev. B} \textbf{\bibinfo{volume}{89}},
  \bibinfo{pages}{245130} (\bibinfo{year}{2014}).

\bibitem[{\citenamefont{Shen et~al.}(2005)\citenamefont{Shen, Ronning, Lu,
  Baumberger, Ingle, Lee, Meevasana, Kohsaka, Azuma, Takano
  et~al.}}]{ShenScience2005}
\bibinfo{author}{\bibfnamefont{K.~M.} \bibnamefont{Shen}},
  \bibinfo{author}{\bibfnamefont{F.}~\bibnamefont{Ronning}},
  \bibinfo{author}{\bibfnamefont{D.~H.} \bibnamefont{Lu}},
  \bibinfo{author}{\bibfnamefont{F.}~\bibnamefont{Baumberger}},
  \bibinfo{author}{\bibfnamefont{N.~J.~C.} \bibnamefont{Ingle}},
  \bibinfo{author}{\bibfnamefont{W.~S.} \bibnamefont{Lee}},
  \bibinfo{author}{\bibfnamefont{W.}~\bibnamefont{Meevasana}},
  \bibinfo{author}{\bibfnamefont{Y.}~\bibnamefont{Kohsaka}},
  \bibinfo{author}{\bibfnamefont{M.}~\bibnamefont{Azuma}},
  \bibinfo{author}{\bibfnamefont{M.}~\bibnamefont{Takano}},
  \bibnamefont{et~al.}, \bibinfo{journal}{Science}
  \textbf{\bibinfo{volume}{307}}, \bibinfo{pages}{901} (\bibinfo{year}{2005}),
  ISSN \bibinfo{issn}{0036-8075}.

\bibitem[{\citenamefont{Amaricci et~al.}(2010)\citenamefont{Amaricci, Camjayi,
  Haule, Kotliar, Tanaskovi\ifmmode~\acute{c}\else \'{c}\fi{}, and
  Dobrosavljevi\ifmmode~\acute{c}\else \'{c}\fi{}}}]{amaricciPRB2010}
\bibinfo{author}{\bibfnamefont{A.}~\bibnamefont{Amaricci}},
  \bibinfo{author}{\bibfnamefont{A.}~\bibnamefont{Camjayi}},
  \bibinfo{author}{\bibfnamefont{K.}~\bibnamefont{Haule}},
  \bibinfo{author}{\bibfnamefont{G.}~\bibnamefont{Kotliar}},
  \bibinfo{author}{\bibfnamefont{D.}~\bibnamefont{Tanaskovi\ifmmode~\acute{c}\else
  \'{c}\fi{}}}, \bibnamefont{and}
  \bibinfo{author}{\bibfnamefont{V.}~\bibnamefont{Dobrosavljevi\ifmmode~\acute{c}\else
  \'{c}\fi{}}}, \bibinfo{journal}{Phys. Rev. B} \textbf{\bibinfo{volume}{82}},
  \bibinfo{pages}{155102} (\bibinfo{year}{2010}).

\end{thebibliography}
\end{document}